\documentclass[12pt]{article}
\usepackage{epsfig,amssymb,graphicx}

\def\beq{\begin{equation}}
\def\eeq{\end{equation}}
\def\bea{\begin{eqnarray}}
\def\eea{\end{eqnarray}}

\newfont{\cursive}{pzcmi at 9pt}
\def\~t{\tilde{t}}

\def\e2phi{\e^{2\Phi}}

\title{\bf{Black holes as local horizons}}
\author{\bf{Alex B. Nielsen} \\
Center for Theoretical Physics,\\
Seoul National University, Seoul 151-747, Korea}

\begin{document}

\maketitle

\begin{abstract}
This talk gives a brief introduction to black hole horizons and their role in black hole thermodynamics. In particular a distinction is made between quasi-locally defined horizons and event horizons.  Currently some new techniques have led to interesting developments and the field seems to be growing in two distinct directions. We will show how thermodynamics can equally well be applied to locally defined horizons and discuss some recent results. The emphasis is on giving simple intuitive pictures and mathematical details are largely omitted.
\end{abstract}

\section{Introduction}

Black holes play a huge role in contemporary physics. Within the past fifteen years or so, a great deal of astrophysical data has accumulated that black holes are physical objects within our universe, ranging from solar-mass sized black holes to the giant objects that inhabit the centres of galaxies and quasars. However, on closer inspection one notices that the astrophysical data only provides evidence for {\it{black hole-like}} spacetimes exterior to the black hole and that current observations are unable to demonstrate that the objects within these regions are ``true" black holes\footnote{This fact has, for example, allowed some speculation on the possibility that the sources of these gravitational fields might be something like gravastars instead of black holes.}. In fact, due to the definition of an event horizon, it would seem virtually impossible to demonstrate that these objects are truly black holes, since no signal will ever reach us from the black hole itself.

This is of course related to the issue of what the definition of a black hole should be. Most lay people have heard of black holes and most physicists think they know what they are. The most common response to the question "what is a black hole?" would probably be "an object whose gravitational field is so strong not even light can escape from it". A slightly more technical response might be "the region inside an event horizon", where the event horizon is understood as the past causal boundary of some region, usually future null infinity. In fact, in most people's minds black holes have become almost synonymous with event horizons. This fact is important to remember as event horizons are by their very definition, non-local structures. In some sense, they `know' about the full global causal structure of space-time, and in particular they know about the future causal structure.

Event horizons are however, not the only way of defining black holes in general relativity. In fact, they are not even the first formal definition appearing in the literature of what it means to be a `black hole'. A number of local, or quasi-local definitions of black holes have been given by various authors and the main aim of this talk is to review some of these ideas and discuss what implications they may have for our understanding of black hole thermodynamics.

The debate about the correct definition may seem like so much semantic quibbling. Clearly, once we have an `intuitive' idea of the physical system we want to describe as a black hole, we can merely search for the most appropriate formal definition. But the definition we choose may have subtle effects on the physical properties we ascribe to the object in question. A non-local definition may eventually lead us to associating these objects with non-local physics. In this way, it seems that local definitions of black holes are more in keeping with the formal local nature of general relativity and quantum field theory. It remains to be seen whether such `localness' will also be a feature of a full theory of quantum gravity.

Since the surprising discovery of black hole thermodynamics a lot of work has gone into discovering what are the microstates responsible for gravitational entropy and what is the end-point of black hole evaporation. In this talk I will describe how the thermodynamic analogy can be applied to local horizons as well as event horizons and that the issue of which horizon best describes the thermal properties of the black hole is not entirely moot.

In order to understand how we came to our present situation we will first discuss some of the history of black holes. We will then discuss various definitions that have been given for black hole horizons and focus on the use of locally defined horizons and their relation to thermodynamics. Finally, we will conclude with some speculative remarks on how this might shed light on various black hole questions.

\section{Brief history}

The idea of a black hole has a long history. The idea of an object so heavy that corpuscular light cannot escape it dates back Reverend John Mitchell in 1783 and Pierre-Simon Laplace in 1795. Newtonian black holes are somewhat different to their relativistic counterparts. While the light particles were able to escape off the surface of the object, they were eventually pulled back down by gravity and this Newtonian picture relies critically on the idea that light particles cannot escape to infinity from the black hole. We thus already see in this some beginning to the global conception of what a black hole should be.

The idea of a black hole in general relativity had a slow start from the original mathematical solution by Karl Schwarzschild 1916, through the work of Oppenheimer and colleagues in the 1930's and Wheeler and colleagues in the 1950's\footnote{A popular account of the history of black hole physics and how they came to be considered physically realistic can be found in \cite{Thorne:book}.}. This era was chiefly concerned with black holes as the endpoint of stellar collapse and it took the community some time to accept that these objects may actually exist.

The first general mathematical definition of what we might heuristically call a black hole was given by Penrose in 1964 \cite{Penrose:1964wq}. This was in terms of what are called trapped surfaces and formed part of his famous singularity theorem.

A trapped surface is a compact, two dimensional spacelike surface for which the forward pointing null normals both have negative expansion. Effectively this means that null rays that are moving away from the surface are forced to move towards one another. Since the surface is a two dimensional spacelike surface, there are two distinct null normal directions. In the following we will think of one of these directions as being ingoing and the other as being outgoing.

In terms of the two vectors tangent to the ingoing and outgoing null geodesics ($n^{a}$ and $l^{a}$ respectively), the expansions $\theta_{n}$ and $\theta_{l}$ can be written as
\beq \theta_{n} = g^{ab}\nabla_{a}n_{b} + l^{a}n^{b}\nabla_{a}n_{b} + n^{a}l^{b}\nabla_{a}n_{b} \eeq
and
\beq \theta_{l} = g^{ab}\nabla_{a}l_{b} + l^{a}n^{b}\nabla_{a}l_{b} + n^{a}l^{b}\nabla_{a}l_{b} \eeq
Each of these expansions only depends on a choice of tangent vector and the metric components at a point in spacetime and so are purely local quantities. To verify that the expansions of both $\theta_{n}$ and $\theta_{l}$ are negative over a compact two surface, requires one to measure their value over a region of spacetime and hence the definition of a trapped surface is actually a quasi-local definition.

On a given time-slicing, the boundary of the region containing trapped surfaces came to be known as a marginally trapped surface. In this case, one of the expansions is zero. This is usually chosen to be the expansion of the outgoing null normal and hence one talks of an marginally outer trapped surface. This marginally outer trapped surface is sometimes called the apparent horizon. 

In 1971 Hawking proposed an alternative definition of a black hole based on the formal idea of an event horizon \cite{Hawking:1971vc}\footnote{The idea of an event horizon had been around since the time of Finkelstein and appears in many works prior to Hawking's formal definition, such as \cite{Israel:1967wq}.}. The event horizon was defined as the past causal boundary of future null infinity and the black hole was defined as the region not contained in the causal past of future null infinity. In practical terms this means that observers inside the black hole region cannot send causal messages to infinity and thus they cannot send causal messages to observers outside the black hole either (since these observers would otherwise be able to just pass the message on to infinity). 

What this meant was that black holes were defined in a global sense. It would never be possible for observers inside the black hole to send causal messages to infinity no matter how long they waited. Whether you were inside or outside the black hole depended on events in the far future of which you could have no knowledge.

\begin{figure}
\includegraphics[scale=0.8]{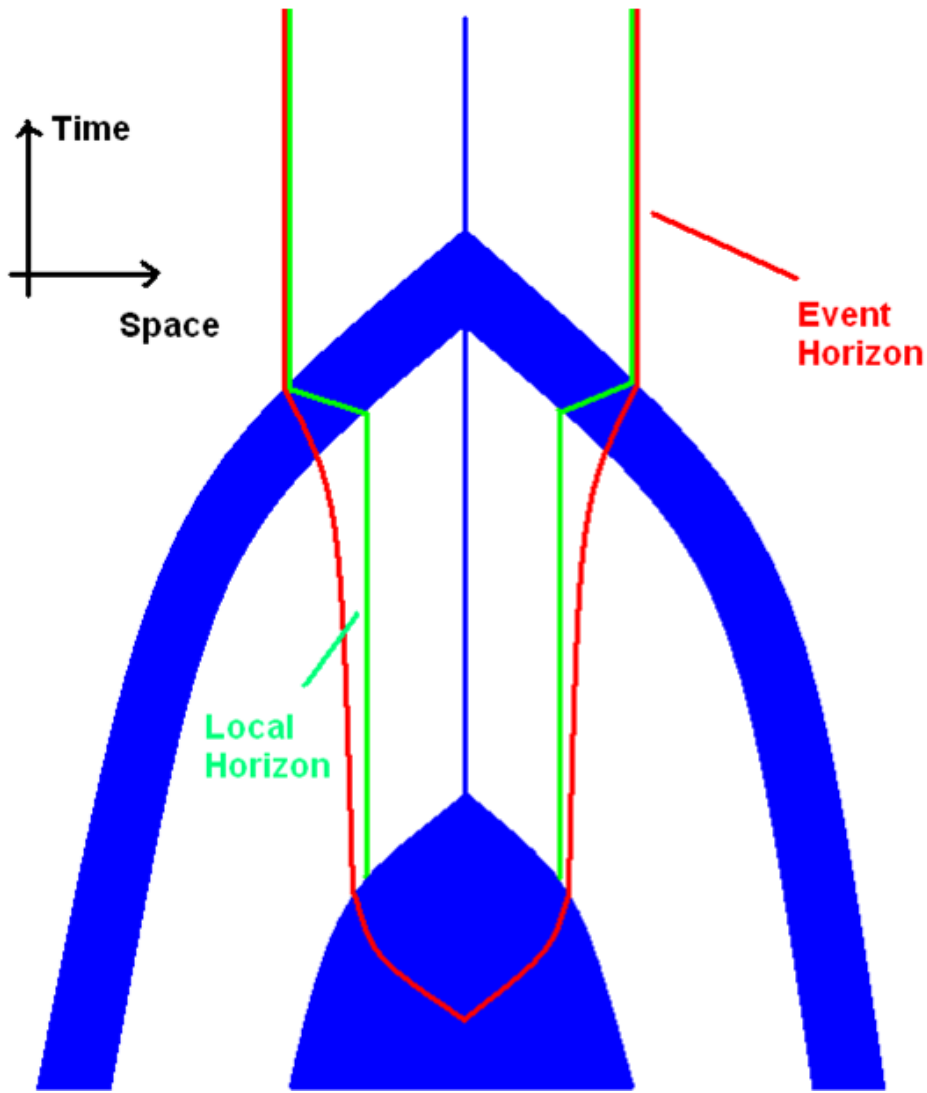}
\label{thickshellcollapse}
\caption{Schematic space-time diagram of a black hole collapse followed by the collapse of a spherical shell onto the black hole (blue shading). Notice that the event horizon (red line) evolves continuously and moves outwards in anticipation of the collapse of the spherical shell whereas the local/apparent horizon (green line) does not respond until it meets the shell itself. Note also that there is a region between the event horizon and the local horizon where light rays can move outwards locally but they will ultimately be trapped by the collapsing shell. The diagram assumes that the future development of the black hole is entirely classical and no further energy is transmitted across the event horizon.}
\end{figure}

It is important to realise that the two notions of apparent horizon and event horizon do not always coincide. In eternally static solutions solutions such as the Schwarzschild solution and the Reissner-Nordstr\"{o}m solution the event horizon is also an apparent horizon. However, in dynamical situations the apparent horizon need not be at the same location as the event horizon. This can be seen clearly in (the colour version of) Fig. \ref{thickshellcollapse}. This means that if one wants to assign an entropy to the horizon of a black hole, or postulate a set of microstates on or near the horizon that give rise to this entropy, one must indicate which horizon one is talking about.

The distinction may be lost when considering globally static toy models, but clearly any physically realistic evaporating black hole spacetime will be dynamic and so in this case and particularly in this case one must choose whether one wants to focus on a globally defined concept, the event horizon, or a locally defined concept such as the apparent horizon.

\begin{figure}
\includegraphics[scale=0.8]{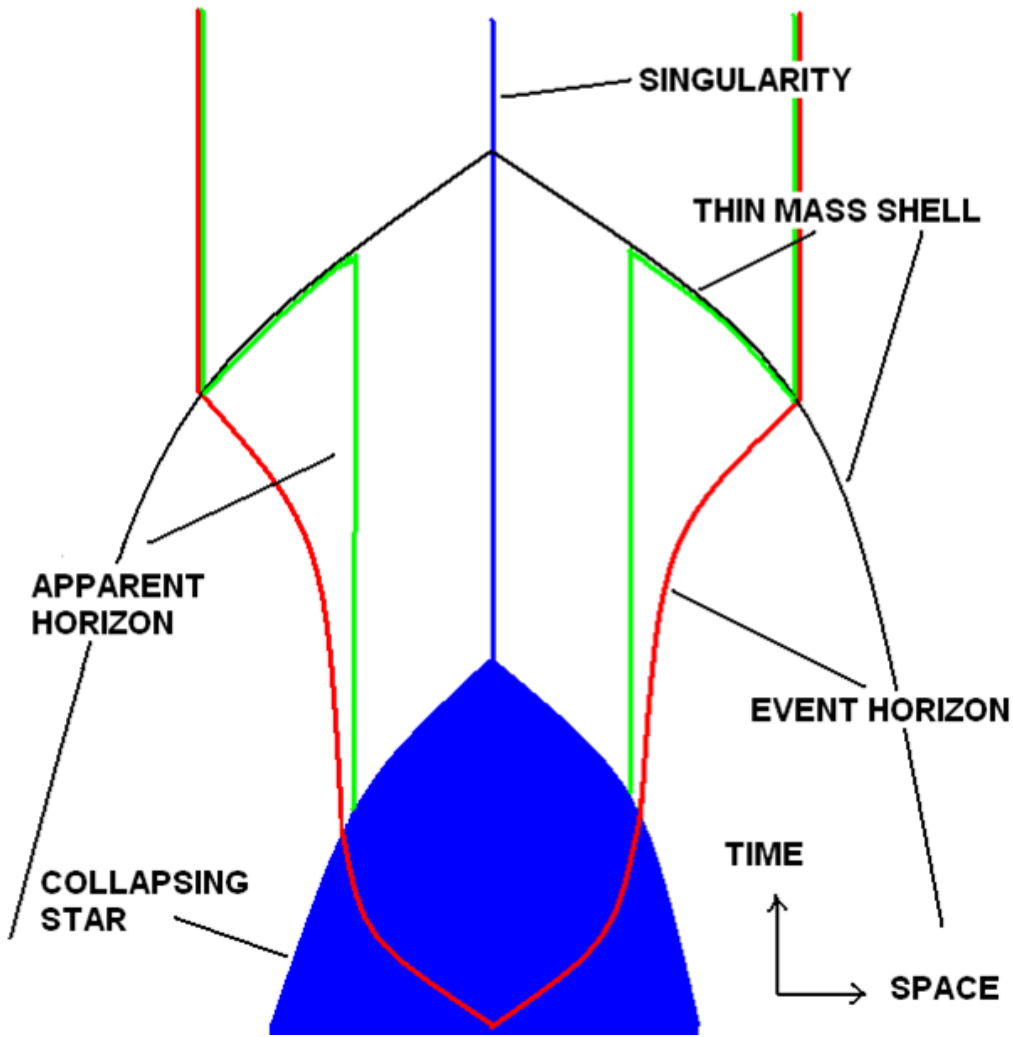}
\label{thinshellcollapse}
\caption{Similar to Fig.\ref{thickshellcollapse} except this time the collapsing shell is a very thin shell. In this case the evolution of the local horizon on a foliation by spacelike hypersurfaces (for example horizontal sections of the diagram) is discontinuous. The area of the outermost horizon jumps at the point where the outer horizon region appears. However, the evolution of the event horizon is still perfectly continuous.}
\end{figure}

There were several reasons given by Hawking (and others) for focusing on event horizons instead of apparent horizons. Chief amongst these was that the evolution of the event horizon is always continuous. This problem is most clearly demonstrated in the case of a very thin shell of matter collapsing onto a pre-existing black hole, see Fig.\ref{thinshellcollapse}. In this case the event horizon evolves continuously, but the apparent horizon does not.

Another related feature of the event horizon is, since it is by definition a null hypersurface, it cannot be multiply intersected by a given spatial hypersurface. As we will see below, locally defined horizons are not always null hypersurfaces. In many cases they are spacelike hypersurfaces and can thus be intersected multiple times by a given spacelike, constant time, slicing.

Perhaps the biggest weakness of apparent horizons and other similarly defined local horizons is their reliance on a particular foliation of the spacetime. For a given foliation of spacetime one can search for compact spacelike two-surfaces. However, not all foliations necessarily give rise to trapped compact surfaces, a point made by Wald and Iyer \cite{Slicing} and not all foliations that give rise to trapped surfaces give rise to trapped surfaces in the same location. This issue is related to a conjecture by Eardley that will be discussed below.

For these reasons apparent horizons were largely ignored in future work by black hole researchers and the concept of an event horizon became synonymous with a black hole in most people's minds. Historically it is important to realise that these definitions were laid down and accepted by the community before anyone started taking seriously the possibility that black holes might have entropy and that they might evaporate by Hawking radiation. 

However, as noted by Hayward \cite{Hayward:1993wb} and Ashtekar and colleagues (see \cite{Ashtekar:2004cn} and references therein), there are also a number of features of event horizons that make them objectionable as the tangible physical boundary of a black hole. Firstly, there are practical considerations for researchers investigating black holes numerically on computers. Locating event horizons in dynamical simulations is notoriously difficult, that is to say time consuming (see for example \cite{Thornburg:2006zb}). In addition, finding event horizons in numerical solutions requires a solution that is stable all the way to `infinity', or at least until it settles down to an approximately stationary state. It is far easier numerically to locate apparent horizons on a given hypersurface and in many cases, use this as a `proxy' for the event horizon.

But this is merely a matter of technical convenience. What is important physically is identifying the mathematical structures that will best allow us to predict the outcome of experiments. Because of the global nature of their definition, classically one cannot detect an event horizon using local measurements. This means that there is no finite experiment one can perform that will demonstrate the physical existence of an event horizon. Such a demonstration would require knowledge of the full future evolution of the universe \emph{all the way out to infinity}. Defining black holes by event horizons means that physics is no longer concerned solely with the elements of the physical world that can be measured and tested with realistic laboratory equipment of a finite size.

Of course, this inability to detect an event horizon only holds classically. But even if one claimed that the event horizon is the structure that is responsible for Hawking radiation, and one was able to detect Hawking radiation, one could still not unequivocally demonstrate that the Hawking radiation was associated with the event horizon, since one would still not be able to prove where the event horizon would be.

Further mischief occurs when we realise that event horizons can occur perfectly well in flat spacetime. Consider a hollow, spherically symmetric shell of matter collapsing spherically symmetrically onto itself. By the shell theorem we know that the metric inside the hollow shell must be that of flat spacetime. But we also know that at some point it will not be possible to send a photon to infinity from within the shell since if we wait too long the escaping photon will reach the collapsing shell after the shell has crossed its Schwarzschild radius and the photon will not be able to escape. Thus there must also be an event horizon inside the collapsing shell, growing out through \emph{flat spacetime}. This leaves open the possibility that there is an event horizon passing through us \emph{at this very moment} due to some exceptionally large shell of matter collapsing down on us from beyond our past light cone. Could it be that my last chance to send a signal to infinity has passed just this very moment and all that I wish to say is now doomed to collapse with me into some future black hole?

Were we to argue that quantum mechanics would endow such event horizons with physical properties we should in principle be able to detect these horizons. The non-existence of these effects would therefore allow us to infer information about events occurring beyond our past light cone. Note also that there would be no local backscatter for such an event horizon in flat space since there is no curvature near the horizon and thus any radiation produced would not have any grey-body factors.

In addition, and perhaps most importantly, there are a number of simple derivations of the Hawking effect that do not require event horizons, indeed depend on local horizons for their derivation \cite{Parikh:1999mf,Visser:2001kq}. Indeed, it would be very strange if the combination of two local theories, namely general relativity and quantum field theory could give rise to non-local particle creation, based purely on event horizons, at the semi-classical level. It's possible that these theories require non-locality for their ultimate consistency and that a fully consistent unification of these two theories does lead to such acausal effects, but developing such a unification is left as an exercise for the reader.

\section{Types of locally defined horizons}

\noindent We turn now to the definitions of local horizons. The two most studied recent definitions for local black hole horizons are those of trapping horizons, due to Hayward \cite{Hayward:1993wb}, and dynamical horizons, due to Ashtekar and collaborators \cite{Ashtekar:2004cn}. (The earlier work by Ashtekar et al. on Isolated Horizons largely refers to a limiting case of a dynamical horizon where the area of the horizon remains constant.)
Both of these definitions are given in terms of the properties of an outgoing null direction $l$ and an ingoing null direction $n$.\bigskip

\noindent A \textbf{future outer trapping horizon} is defined by Hayward
\cite{Hayward:1993wb} as the closure of a three-surface which is
foliated by marginal surfaces, for which $\theta_{l}=0$,
and which, in addition, satisfies
\begin{enumerate}
\item[\it{i}.] $\theta_{n}<0$ (to distinguish between white holes
and black holes).
\item[\it{ii}.] ${\cal{L}}_{n}\theta_{l}<0$ (to distinguish
between inner and outer horizons of, for example, the non-extremal Reissner-Nordstr\"{o}m solution).
\end{enumerate}\bigskip
A \textbf{dynamical horizon} $H$ is
defined by Ashtekar and Krishnan \cite{Ashtekar:2004cn} to have
the following properties;

\begin{enumerate}
\item[\it{i}.] \emph{H} is a three-dimensional spacelike
hypersurface that can be foliated by closed, spacelike
2-surfaces.
\item[\it{ii}.] The expansion of one null normal to the foliations
$n^{a}$ is negative $\theta_{n}<0$.
\item[\it{iii}.] The expansion of the other null normal $l^{a}$ is
zero $\theta_{l}=0$.
\end{enumerate}
These definitions are similar to the definition of a marginally trapped surface in that they rely crucially on the condition $\theta_{l} = 0$. The definitions are also quite similar to each other. Most dynamical horizons are trapping horizons, although trapping horizons can be timelike, null or spacelike hypersurfaces. However, the definitions are subtly different in intent in that a dynamical horizon does not refer to any behaviour off of the horizon, while condition ($ii$) for a future outer trapping horizon refers to the behaviour of $\theta_{l}$ as one moves across the horizon.

As mentioned earlier, one of the chief criticisms of locally defined horizons has been that they are foliation dependent. Different foliations of the spacetime may give rise to different local horizons and some foliations may give rise to no local horizons at all. The definitions of trapping horizons and dynamical horizons sidestep this issue somewhat by not referring to behaviour on a given spacelike hypersurface but the question of their uniqueness still remains \cite{Ashtekar:2005ez}.

In fact, there is an intriguing conjecture due to Eardley \cite{Eardley:1997hk} that relates local horizons in their marginally trapped surfaces guise to event horizons.

Eardley showed how one could deform a given smooth marginally trapped surface by choosing a new foliation in such as way as to move the location of the horizon slightly outwards. Since one can show that marginally trapped surfaces generally occur inside event horizons, Eardley conjectured that one could continue this process of `moving' the horizon outwards until one reached the event horizon.

This would seem to give a central role back to the event horizon even if one initially focuses on local horizons since, if the conjecture holds, the boundary of the region where one can find marginally trapped surfaces is precisely the region bounded by the event horizon.

However a key assumption of the proof that marginally trapped surfaces always lie inside event horizons is that of the Null Energy Condition. This condition can be written as
\beq T_{ab}l^{a}l^{b} \geq 0 \eeq
or, for a perfect fluid
\beq \rho + p \geq 0 \eeq
While this is a perfectly reasonable assumption for all known classical matter, it is well-known to be violated by quantum fields during Hawking radiation. Thus, in precisely the situations one would want to consider in black hole evaporation, the conjecture is unlikely to hold and the event horizon has no real role.

One way to sidestep these issues is to focus purely on the properties of a given trapping or dynamical horizon and not to concern oneself too much with how the horizon itself came about. This is the approach largely adopted by Hayward and Ashtekar and is the approach we will take for the remainder of the talk.

\section{Local horizons and thermodynamics}

We turn now to the issue of black hole thermodynamics. The fact that black hole dynamics can be cast in an equivalent form to the usual laws of thermodynamics has long been the subject of much theoretical interest. The laws of black hole thermodynamics were first demonstrated for event horizons. The initial proofs relied on both the global Killing vector field of a globally stationary spacetime and asymptotic flatness to normalise the surface gravity (of the Killing horizon) and to define the mass of the black hole as the asymptotic ADM mass.

That the laws of thermodynamics hold for event horizons would seem to suggest that any investigation into the thermodynamic properties of black holes should focus on event horizons. But it is now clear that the same laws can be derived for locally defined horizons too. For our purposes here we will only consider the rather simple proof given by Hayward \cite{Hayward:1993wb} of the second law. Since the zeroth law is not expected to hold for dynamical black holes and the first law really relates to the conservation of energy the interested reader is referred to \cite{Ashtekar:2004cn} for these cases.

The tangent vector field of a trapping horizon can be written as
\beq r^{a} = \alpha l^{a} + \beta n^{a} \eeq
Where will we choose for convenience $\alpha > 0$ and $n^{a}l_{a} = -1$. With this choice we immediately see that the norm of the horizons tangent vector satisfies
\beq r^{a}r_{a} = -2\alpha\beta \eeq
and thus the horizon can be timelike, null of spacelike depending on the sign of $\beta$. The change in the area $A$ as we move along the horizon can be computed by
\beq {\cal{L}}_{r}A = \alpha {\cal{L}}_{l}A + \beta {\cal{L}}_{n}A \eeq
We can relate the change in the area to the expansion via the well-known equation
\beq \theta_{l}A = {\cal{L}}_{l}A \eeq
which holds for any vector, not just the outgoing radial null geodesic $l^{a}$. Thus, since $\theta_{l}$ is assumed zero on the horizon we have
\beq {\cal{L}}_{r}A = \beta\theta_{n}A \eeq
Since $\theta_{n}$ is assumed negative, we see that a spacelike horizon has an increasing area and a timelike horizon has a decreasing area. We can also relate the value of $\beta$ to the value of $\alpha$ by noticing that we require
\beq {\cal{L}}_{r}\theta_{l} = \alpha {\cal{L}}_{l}\theta_{l} + \beta {\cal{L}}_{n}\theta_{l} = 0 \eeq
If $l^{a}$ is hypersurface orthogonal (not necessarily hypersurface orthogonal to the horizon but hypersurface orthogonal to some surface) then by the Raychaudhuri equation we can write
\beq  {\cal{L}}_{r}\theta_{l} = - \sigma^{2} - R_{ab}l^{a}l^{b} \eeq
and hence we have
\beq \beta = \alpha \frac{\left( \sigma^{2} + R_{ab}l^{a}l^{b}\right)}{{\cal{L}}_{n}\theta_{l}} \eeq
which gives
\beq {\cal{L}}_{r}A =  \frac{\alpha A \theta_{n}\left( \sigma^{2} + R_{ab}l^{a}l^{b}\right)}{{\cal{L}}_{n}\theta_{l}} \eeq
Via the Einstein equations we can cast this in the form
\beq {\cal{L}}_{r}A =  \frac{\alpha A \theta_{n}\left( \sigma^{2} + 8\pi G T_{ab}l^{a}l^{b}\right)}{{\cal{L}}_{n}\theta_{l}} \eeq
Since $\alpha > 0$, $\theta_{n} < 0$ and ${\cal{L}}_{n}\theta_{l} < 0$ for all trapping horizons, we see that the second law of thermodynamics holds as long as the Null Energy Condition is satisfied and would fail only in situations where the NEC does not hold as indeed is the expected case with Hawking radiation. That the second law holds equally well for trapping horizons as for event horizons is the main result of interest.

On a given spatial hypersurface, the area of the trapping horizon and the area of the event horizon may not be the same. This tells us that any attempt to incorporate the area of black holes into a generalised entropy function, or identify the microstates responsible for that entropy must inevitably face the question of which horizon is being referred to.

\section{Local horizons and information loss}

We turn finally to a brief discussion of the role that horizons could play in the evaporation of black holes. While the discussion is brief and heuristic it will hopefully indicate what role the choice of horizon may play in resolving certain problems.

\begin{figure}
\includegraphics[scale=0.8]{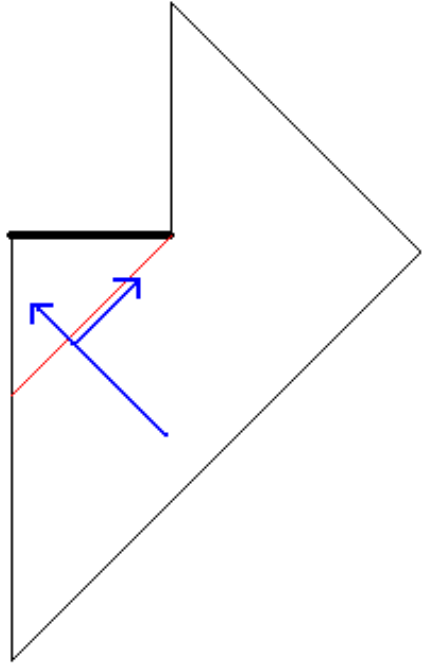}
\label{complement_bh}
\caption{Penrose diagram of black hole evaporation according to the black hole complementarity picture. Infalling matter is somehow `duplicated' at the horizon such that observers both inside and outside the horizon have access to the information it contained. Unitarity is preserved for both observers and only violated for a `superobserver' who can see the whole spacetime, something that precludes them from being a true observer.}
\end{figure}

Firstly we consider the black hole complementarity picture first elucidated in \cite{Susskind:1993if} and displayed in Fig. \ref{complement_bh}. In this picture one assumes that there is a true event horizon and that unitarity is preserved. The argument runs, that these two are only compatible if observers on the outside and observers on the inside see different physics. From the point of view of an observer who remains outside the horizon, the horizon is represented by a hot membrane (in the sense of the stretched horizon) that strongly interacts with all infalling matter, eventually re-emitting the matter as Hawking radiation. However, complemetary to this, from the point of view of observers that fall over the horizon with the infalling matter, nothing unusual occurs at the horizon and the infalling matter carries on its path towards the central singularity.

This picture appears to be entirely consistent in that there is no experiment that could be performed to inform either of the causally disconnected observers of what the other observer has seen. There is no way to experimentally verify the duplication of matter. This feature seems to depend on the `hot membrane' being associated with the event horizon and thus the picture seems to rely from the outset on a non-local description. In fact, the whole rationale for considering what to some may seem an extreme solution is predicated on considering event horizons in the context of unitary evolution. As we will see below, it may not be necessary to go this far if one is willing to consider spacetimes without true event horizons.

The picture also raises a number of questions. For example, would an observer that initially follows a path that keeps them out of the black hole and observes matter being thermalised by the hot membrane suddenly see the membrane disappear if they choose to allow themselves to fall freely into the black hole? And how would the picture look if we consider, as we did earlier, an event horizon forming in an empty region of space, with a shell of matter collapsing down from beyond the past light cone? Would an observer in the flat region outside the event horizon still `see' a hot membrane forming on the event horizon?

\begin{figure}
\includegraphics[scale=0.8]{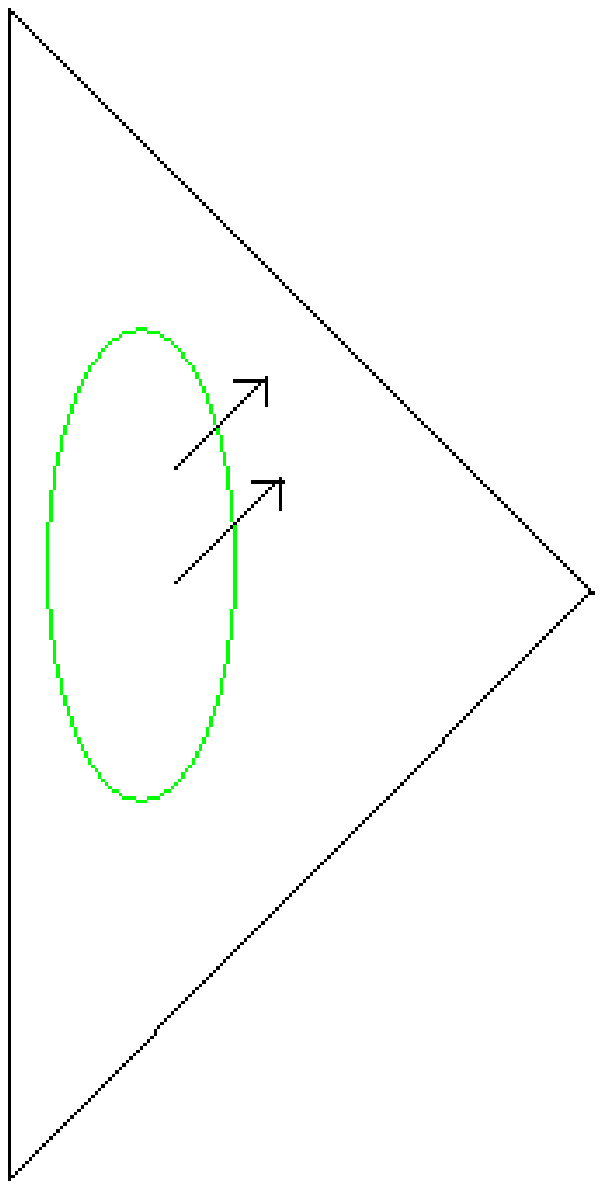}
\label{hay_bh}
\caption{Penrose diagram of black hole evaporation according to Hayward. The central singularity no longer exists, the centre of the spacetime is timelike and there is no event horizon. Only the trapping horizon (green line) exists and information will be able to travel freely out of the black hole when the trapping horizon is timelike.}
\end{figure}

A more straightforward approach is that of Hayward \cite{Hayward:2005gi} and Fig. \ref{hay_bh}. Hayward argues that the details of what is occuring near the singularity should not make a difference to the general picture of Hawking radiation and also that one should think of a black holes in terms of trapping horizons rather than event horizons. To this end Hayward proposes a model that contains a black hole with two trapping horizons, an inner and an outer trapping horizon and a timelike $r=0$ region. His spacetime contains no event horizon and Hayward argues that information that fell into the black hole can be returned to the exterior region whilst the trapping horizon is timelike and evaporating.

The original spacelike singularity that appears in the semi-classical picture has now been modified to be everywhere timelike and there is no real mechanism proposed to account for the production of Hawking radiation and how it should be correlated to infalling material. Rather Hayward makes the point that the existence of an event horizon is not strictly necessary to understand black hole formation and evaporation, much as we have here in this talk.   

\begin{figure}
\includegraphics[scale=0.8]{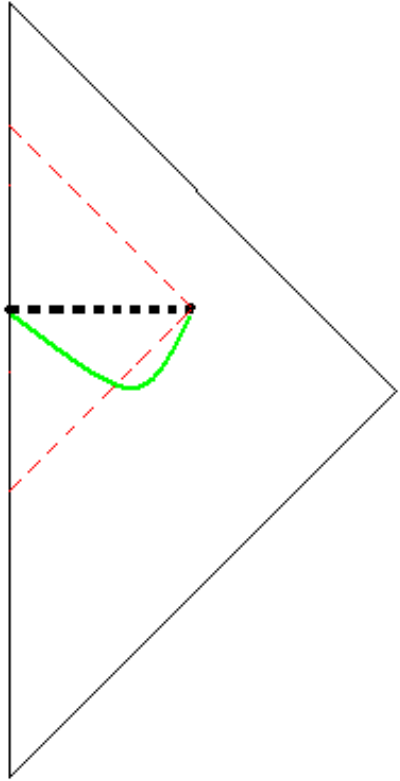}
\label{ash_bh}
\caption{Penrose diagram of black hole evaporation according to Ashtekar and Bojowald. The central singularity is resolved by quantum gravity effects and it is possible to continue the `spacetime' through the formally singular region (thick dotted line). There is no true event horizon but there is a region within which one cannot avoid falling through the quantum spacetime region indicated by the lower dotted red line.}
\end{figure}

Another possibility is that of Ashtekar and Bojowald \cite{Ashtekar:2005cj} and Fig. \ref{ash_bh}. Based on calculations using the loop quantum gravity formalism Ashtekar and Bojowald claim that it is possible to extend spacetime through the central singular region. Of course the spacetime is not smooth all the way through this extension but it is claimed that a classical smooth manifold is eventually reached on the other side. This means that one can extend the usual semi-classical picture of black hole evaporation in the way depicted in the figure. This avoids the extreme modification made in the Hayward picture to the central region since the formally singular region is still spacelike, only now it does not form a true boundary to the spacetime and geodesic curves do not necessarily end when they intersect it. The geometry is also not smooth here, something that one might reasonably expect from a theory of quantum gravity.

This means that there is no true event horizon in the spacetime, only a dynamical/trapping horizon, and the event horizon is instead replaced by a `quantum horizon' which denotes the boundary of the region within which one cannot avoid meeting the formally singular area. While the model appears to have some attractive features, it is fair to say that more work remains to be done to see if it is truly viable. For example, one should calculate whether correlations between matter that fell in to the black hole and matter that stayed outside can be sustained whilst the infalling matter crosses the central region.

\section{Conclusions}

The issue of what is the most useful description of a black hole depends on the context. In many astrophysical situations one may be happy thinking of a black hole as merely "the end-point of gravitational collapse of a sufficiently massive object". This definition may not preclude several exotica that one might not be willing to call a black hole in a theoretical context. In this case it would be perfectly reasonable to invoke the event horizon as a defining boundary. This is undoubtedly a perfectly reasonable, self-consistent definition.

However, one should remember that this definition is not the only possible definition and may not be the most useful when considering semi-classical effects such as Hawking radiation. In fact its causal nature may eventually lead to one entertaining causally separated views of reality such as are seen in the black hole complementarity picture. As we have seen above, it is possible to consider spacetimes that do not contain event horizons but act essentially like black holes. That fact that one will probably never be able to prove that a true event horizon exists in nature may ultimately force us to consider this option. Moreover, if one wants to assign an entropy to the event horizon merely because "information is eternally lost behind it" one has a consistency issue if one simultaneously wants to argue that the information is eventually returned to the universe.

Adopting quasi-local definitions of horizons doesn't necessarily solve all of the above problems. Since marginally trapped surfaces are not unique, trapping and dynamical horizons may not be unique either. Whether one wants to consider the outermost possible local horizon or some other restriction will need to be carefully studied. Spacelike horizons may also intersect a given constant time surface multiple times giving the sense of multiple horizons.

As we have seen above, these issues are far from settled and may have to wait until we understand better what exactly occurs near the centre of the black hole, where classically one would expect a singularity to form. Until then we should be wary of assigning too much physical significance to what are essentially just definitions.

\section{Acknowledgements}

It is a pleasure to thank Ewan Stewart and Dong-han Yeom for useful comments and their hospitality at KAIST.


\begin{thebibliography}{999}

\bibitem{Thorne:book}
  K.~S.~Thorne
  ``Black holes and time warps: Einstein's outrageous legacy,''
  W.~W.~Norton and Co. Inc (1994)

\bibitem{Penrose:1964wq}
  R.~Penrose,
  Phys.\ Rev.\ Lett.\  {\bf 14} (1965) 57.

\bibitem{Israel:1967wq}
  W.~Israel,
  Phys.\ Rev.\  {\bf 164} (1967) 1776.

\bibitem{Hawking:1971vc}
  S.~W.~Hawking,
  Commun.\ Math.\ Phys.\  {\bf 25} (1972) 152.

\bibitem{Slicing}
R. M. Wald and V. Iyer, ``Trapped surfaces in the Schwarzschild
geometry and cosmic censorship'', Phys.\ Rev.\ D {\bf 44} (1991)
R3719--R3722.

\bibitem{Hayward:1993wb}
  S.~A.~Hayward,
  Phys.\ Rev.\  D {\bf 49} (1994) 6467.

\bibitem{Ashtekar:2004cn}
  A.~Ashtekar and B.~Krishnan,
  Living Rev.\ Rel.\  {\bf 7} (2004) 10
  [arXiv:gr-qc/0407042].

\bibitem{Thornburg:2006zb}
  J.~Thornburg,
  Living Rev.\ Rel.\  {\bf 10} (2007) 3
  [arXiv:gr-qc/0512169].

\bibitem{Parikh:1999mf}
  M.~K.~Parikh and F.~Wilczek,
  Phys.\ Rev.\ Lett.\  {\bf 85} (2000) 5042
  [arXiv:hep-th/9907001].

\bibitem{Visser:2001kq}
  M.~Visser,
  Int.\ J.\ Mod.\ Phys.\  D {\bf 12} (2003) 649
  [arXiv:hep-th/0106111].

\bibitem{Ashtekar:2005ez}
  A.~Ashtekar and G.~J.~Galloway,
  Adv.\ Theor.\ Math.\ Phys.\  {\bf 9} (2005) 1
  [arXiv:gr-qc/0503109].

\bibitem{Eardley:1997hk}
  D.~M.~Eardley,
  Phys.\ Rev.\  D {\bf 57} (1998) 2299
  [arXiv:gr-qc/9703027].

\bibitem{Susskind:1993if}
  L.~Susskind, L.~Thorlacius and J.~Uglum,
  Phys.\ Rev.\  D {\bf 48} (1993) 3743
  [arXiv:hep-th/9306069].

\bibitem{Hayward:2005gi}
  S.~A.~Hayward,
  Phys.\ Rev.\ Lett.\  {\bf 96} (2006) 031103
  [arXiv:gr-qc/0506126].

\bibitem{Ashtekar:2005cj}
  A.~Ashtekar and M.~Bojowald,
  Class.\ Quant.\ Grav.\  {\bf 22} (2005) 3349
  [arXiv:gr-qc/0504029].

\end{thebibliography}
\end{document}